\documentclass[journal]{IEEEtran}

\usepackage{xcolor,colortbl}
\usepackage{graphicx}
\usepackage{color}
\usepackage{tabularx}
\usepackage{subcaption}
\usepackage{soul}
\usepackage{boldline}
\usepackage{multirow}
\usepackage{booktabs}
\usepackage{url}
\usepackage{amssymb, amsmath, bm}
\usepackage{comment}

\begin{document}

\title{Use of speaker recognition approaches for learning and evaluating embedding representations of musical instrument sounds}

\author{Xuan Shi, Erica Cooper, \IEEEmembership{Member, IEEE}, and Junichi Yamagishi, \IEEEmembership{Senior Member, IEEE}
\thanks{Manuscript received July 22, 2021; revised November 8, 2021; revised December 22, 2021; accepted December 22, 2021. This work was supported in part by JST CREST Grants (JPMJCR18A6 and JPMJCR20D3),  JST AIP Challenge Grant, and MEXT KAKENHI Grants (18H04112, 21H04906, 21K11951). The associate editor coordinating the review of this manuscript was Prof. Stefan Bilbao. \textit{(Corresponding author: Xuan Shi.)}}
\thanks{Xuan Shi is with the University of Southern California, Los Angeles, CA 90089 USA (e-mail: xuanshi@usc.edu)}
\thanks{Erica Cooper and Junichi Yamagishi are with the National Institute of Informatics,  Tokyo 101-8340, Japan (e-mail: \{ecooper, jyamagis\}@nii.ac.jp).}}

\markboth{IEEE/ACM Transactions on Audio, Speech, and Language Processing,~Vol.~XX, No.~XX, XX~2021}
{Shell \MakeLowercase{\textit{et al.}}: Bare Demo of IEEEtran.cls for IEEE Journals}
\maketitle

\begin{abstract}


Constructing an embedding space for musical instrument sounds that can meaningfully represent new and unseen instruments is important for downstream music generation tasks such as multi-instrument synthesis and timbre transfer.  The framework of Automatic Speaker Verification (ASV) provides us with architectures and evaluation methodologies for verifying the identities of unseen speakers, and these can be repurposed for the task of learning and evaluating a musical instrument sound embedding space that can support unseen instruments.  Borrowing from state-of-the-art ASV techniques, we construct a musical instrument recognition model that uses a SincNet front-end, a ResNet architecture, and an angular softmax objective function.  Experiments on the NSynth and RWC datasets show our model's effectiveness in terms of equal error rate (EER) for unseen instruments, and ablation studies show the importance of data augmentation and the angular softmax objective.  Experiments also show the benefit of using a CQT-based filterbank for initializing SincNet over a Mel filterbank initialization.  Further complementary analysis of the learned embedding space is conducted with t-SNE visualizations and probing classification tasks, which show that including instrument family labels as a multi-task learning target can help to regularize the embedding space and incorporate useful structure, and that meaningful information such as playing style, which was not included during training, is contained in the embeddings of unseen instruments.

\end{abstract}


\IEEEpeerreviewmaketitle

\section{Introduction}

Multi-instrument audio synthesis including timbre-style transfer is an actively-researched audio generation task in which we disentangle instrument timbre and music content, control and manipulate the timbre, and generate high-fidelity natural-sounding audio signals. Inputs may be either audio signals of a different instrument or discrete representations such as MIDI. Several models for the former audio-based timbre transfer have been proposed \cite{oord2016wavenet,huang2018timbretron,engel2019gansynth,engel2020ddsp,kim2019neural,mor2018universal}. 
There have also been attempts to generate audio from discrete representations, such as \cite{wang2019performancenet} and  \cite{cooper21midi}. 

An important component in these kinds of multi-instrument audio synthesis is the representation used for instrument types. A very simple representation can be one-hot vectors corresponding to instruments included in a training dataset. But this does not provide us the freedom to use unseen or new instruments that are not included in the training dataset. Another approach is to use a large database that includes monophonic sounds of many different types of instruments and obtain instrument embeddings or latent vectors through a neural encoder \cite{kim2019neural}. This neural encoder is typically simpler than ones used in various music retrieval tasks such as audio identification \cite{wang2003industrial}, audio matching \cite{MuellerKC05_ChromaFeatures_ISMIR}, version identification \cite{SerraGHS08_CoverSong_IEEE-TASLP}, and Jukebox embeddings \cite{castellon2021codified} in terms of neural architectures. On the other hand, there are also problems specific to the instrument encoder used for multi-instrument audio synthesis. For instance, since the embedding or latent vectors corresponding to unseen or new instruments may be specified at inference time, it is critical to design a neural encoder that provides appropriate representations even for such unseen instruments and to evaluate the appropriateness of the instrument embedding vectors obtained from audio of the unseen instruments.

How can we perform such evaluation and analysis of instrument embedding vectors obtained from audio of unseen instruments? In this paper, we propose to adopt evaluation frameworks used in the Automatic Speaker Verification (ASV) field to answer this scientific question. Speaker verification is a task where a claimed speaker identity of an input speech sample is judged to be the same as or different from its template obtained during a registration process called ``enrollment.'' In modern speaker verification systems using neural networks, the templates are typically embedding vectors obtained from a neural speaker encoder, and its distance to the embedding of the input speech is measured to decide whether the input speech's ID is the same as that of the enrolled speech data. Speakers to be verified are normally unseen and their data is not included in the training dataset for the speaker encoder. Hence, we can hypothesize that this evaluation manner and metric used in the speaker verification task is suitable for evaluating the instrument embedding vectors obtained from audio of unseen instruments. Performance is normally measured using Equal Error Rate (EER) and/or Decision Cost Function (DCF) \cite{martin1997det}. For instrument embeddings, the former, EER, is an appropriate choice  since DCF is a metric specific to biometric applications where we need to consider multiple different use scenarios with different tradeoffs of security and convenience. When we have multiple speaker encoders, their EER differences can be further analyzed through a statistical significance test on EERs \cite{bengio2004statistical}. Speaker embedding vectors can also be analyzed via shallow probing tasks \cite{Raj19probing}. Using these evaluation frameworks for ASV, it would be possible to assess multiple instrument encoders without audio generation and analyze their tendencies. 

As described earlier, a database for training the instrument encoder is typically one including monophonic sounds of many different types of instruments and therefore, thanks to similarities of the task and audio signals, we can also utilize speaker recognition models directly. Therefore, in addition to the evaluation methodologies, we introduce several advanced neural architectures that are frequently used for ASV. More specifically, we introduced SincNet \cite{Ravanelli18SincNet} for feature extraction from raw waveforms, ResNet \cite{He16ResNet} as the main body, learnable dictionary encoding (LDE) \cite{Cai18LDE} for aggregation of audio signals of varying lengths, and angular softmax \cite{Huang18ASoftmax} for class discriminative training. All of these have been reported to perform well on multiple ASV benchmark datasets \cite{Ravanelli18SincNet,chung2018voxceleb2,cai2018exploring,Huang18ASoftmax} and they are expected to make our evaluation and analysis of the instrument embedding vectors obtained from unseen instruments meaningful. Instrument encoders using various combinations of these techniques are trained on two popular monophonic instrument databases, NSynth \cite{nsynth2017} and RWC \cite{Goto03RWC}, and they are used to extract embedding vectors of unseen instruments.

This paper is structured as follows. Section \ref{sec:related-work} overviews multi-instrument audio synthesis and musical instrument recognition. Section \ref{sec:unseen-instrument} describes evaluation frameworks for embedding vectors of unseen vectors including statistical significance tests. Section \ref{sec:model} describes the neural network architectures. Section \ref{sec:exp} shows experimental conditions and results, and our findings are summarized in Section \ref{sec:conclusion}. 

\section{Related Work}
\label{sec:related-work}

In this section, we review the background of this paper in the following fields: multi-instrument audio synthesis and musical instrument recognition.

\subsection{Multi-instrument Audio Synthesis}

Multi-instrument audio synthesis involves generating an instrument performance with the timbre of various target instruments given the pitch, dynamics, and other factors.  From the perspective of input, there are generally two paradigms for multi-instrument audio synthesis: audio input and symbolic input. The former one is also known as timbre transfer. 
To transfer the timbre of the input audio, many generative models are conditioned with the timbre or relevant hidden features, which are extracted in an unsupervised manner, to generate target audio.  Engel \MakeLowercase{\textit{et al.}} adopted WaveNet-style Autoencoders \cite{oord2016wavenet} in \cite{nsynth2017} and Generative Adversarial Networks (GAN) \cite{goodfellow2014generative} in \cite{engel2019gansynth} to capture the timbre of the target instrument and to apply the target timbre through time to generate audio in an autoregressive or parallel sampling method.  Similarly, DDSP \cite{engel2020ddsp} extracted hidden features $z$ with the encoder and synthesised high-fidelity audio by the decoder in combination with a harmonic oscillator and filter. In the application of timbre transfer, DDSP successfully transferred timbre between the singing voice and violin. 

Different from the studies mentioned above, using a symbolic MIDI-derived input representation instead of audio input, Kim \MakeLowercase{\textit{et al.}} proposed Mel2Mel \cite{kim2019neural}: a MIDI piano roll and ad-hoc instrument embedding, learned with a FiLM layer from one-hot instrument labels, are given to generate a Mel spectrum, followed by a WaveNet-based synthesizer to generate the audio of the target instrument.  PerformanceNet \cite{wang2019performancenet} uses an architecture made up of U-net and multi-band convolution blocks to convert MIDI piano rolls into acoustic features, which are then converted into waveforms using the Griffin-Lim algorithm.  In \cite{cooper21midi}, the text-to-speech synthesis architecture Tacotron2 \cite{shen2018natural} and the Neural Source-Filter waveform model \cite{wang2019neural} were adapted to accept polyphonic MIDI piano roll input and trained on piano performance data with aligned MIDI transcriptions to synthesize piano music.  The  MIDI representation can also be used to train models to compose new songs in the symbolic domain, and then synthesize them as waveforms, as in \cite{manzelli2018conditioning, huang2019music}.

\subsection{Musical Instrument Recognition and Relevant Topics}
\label{sec:related_work-musical_instrument_recognition}
A great number of previous studies have focused on instrument recognition from single notes and solo recordings of pieces. 
Fuhrmann \cite{fuhrmann2012phd} comprehensively reviewed various machine learning based musical instrument recognition methods on the eve of deep learning's surge. Readers may refer to \cite{fuhrmann2012phd} for details of conventional musical instrument approaches.

In recent years, deep neural networks (DNN) have generally dominated the field of musical instrument recognition. 
Taenzer \MakeLowercase{\textit{et al.}} \cite{Taenzer19instrfml} explored the influence of different data pre-processing and augmentation methods on the generalization ability of CNNs in western classical instrument family recognition tasks.  Similarly, Ramires \MakeLowercase{\textit{et al.}} \cite{ramires19dataaug} applied various audio effects to musical instrument sound to evaluate the robustness of an instrument classification model and boost its performance. With the help of data augmentation, their proposed method obtained an F1 score of $74.73$ on the NSynth database, while the F1 score of the baseline without augmentation was $73.78$.  Recently, Zeghidour \MakeLowercase{\textit{et al.}} \cite{zeghidour21leaf} devised a new kind of learnable front-end (LEAF) for audio classification. LEAF achieves an F1 score of $72.0$ on the NSynth database, demonstrating its capability in musical instrument recognition.  Concurrent to our work, \cite{ghosh2021deep} pretrained an advanced backbone model on a subset of AudioSet \cite{gemmeke2017audioset} and fine-tuned it on specific databases for downstream tasks. However, for the musical instrument recognition task, the randomly initialized model slightly outperforms their proposed pretrained model.

In addition to monophonic musical instrument recognition, recognizing musical instruments in the polyphonic scenario has also attracted interest from many researchers.  Huang \MakeLowercase{\textit{et al.}} \cite{hung2018frame} regarded polyphonic instrument recognition as a multi-class prediction for each frame.  Han \MakeLowercase{\textit{et al.}} \cite{han2016deep} used a sliding window to predict instrument categories and aggregated the window-level results to predict the predominant instrument in multi-instrument audio.  

\section{Evaluation of instrument embedding vectors obtained from audio of unseen instruments}
\label{sec:unseen-instrument}
\subsection{Procedures to evaluate the embedding vectors obtained from audio of unseen instruments}

For the evaluation of instrument embedding vectors obtained from audio of unseen instruments, typical evaluation metrics for identification tasks such as Micro F1-score, Macro F1-score, and confusion metrics cannot be used. We need to rely on other evaluation procedures and metrics.  Here we describe an evaluation methodology using EERs as a proxy for assessing representations of the instrument embedding vectors obtained from audio of unseen instruments. 

First, an evaluation set is assumed to contain audio files of unseen instruments, and samples are further divided into two sets. Using a trained instrument encoder, the first set is used for extracting embedding vectors for enrollment, and the second set is used for measuring similarity to the embedding vector of the same unseen instrument, and dissimilarity to embedding vectors of other unseen instruments included in the test set. The final performance is assessed by computing EERs. EER is a common evaluation metric for verification tasks. The value of EER indicates the point at which the proportion of false acceptances is equal to the proportion of false rejections. The lower the EER value, the more dissimilar and discriminative the embedding representation for an instrument is compared to other unseen instruments. When a sufficiently large number of instruments are included in the test set, this metric indicates whether an instrument encoder can extract appropriate representations for unseen instruments. 

\subsection{Statistical Significance Analysis}

When we compare EERs of different systems, it is also critical to check the statistical significance of the differences.  For the open-set case of verifying instrument categories, a pair-wise statistical significance analysis of EERs can be conducted using the methodology proposed in \cite{bengio2004statistical}. 

For a pair of models $(A, B)$ in the comparison, a $z$ value is computed using 
\begin{equation}
z=\frac{2|\text{EER}_A - \text{EER}_B|}{\sqrt{\Big[\text{EER}_A (1-\text{EER}_A)+\text{EER}_B (1-\text{EER}_B)\Big]\frac{N_{\text{target}} + N_{\text{non-target}}}{N_{\text{target}} N_{\text{non-target}}}}},
\end{equation}
where $N_{\text{target}}$ and $N_{\text{non-target}}$ denote the number of evaluated target and non-target instrument trials, respectively. The $z$ value is then compared with a threshold value $Z_{\alpha/2}$ decided by a significance level (e.g.\ $\alpha=0.05$) with Holm-Bonferroni correction. If $z\geq Z_{\alpha/2}$, a difference between $A$ and $B$ is supported as a statistically significant one.

\begin{figure}
\centering 
\includegraphics[width=0.5\textwidth]{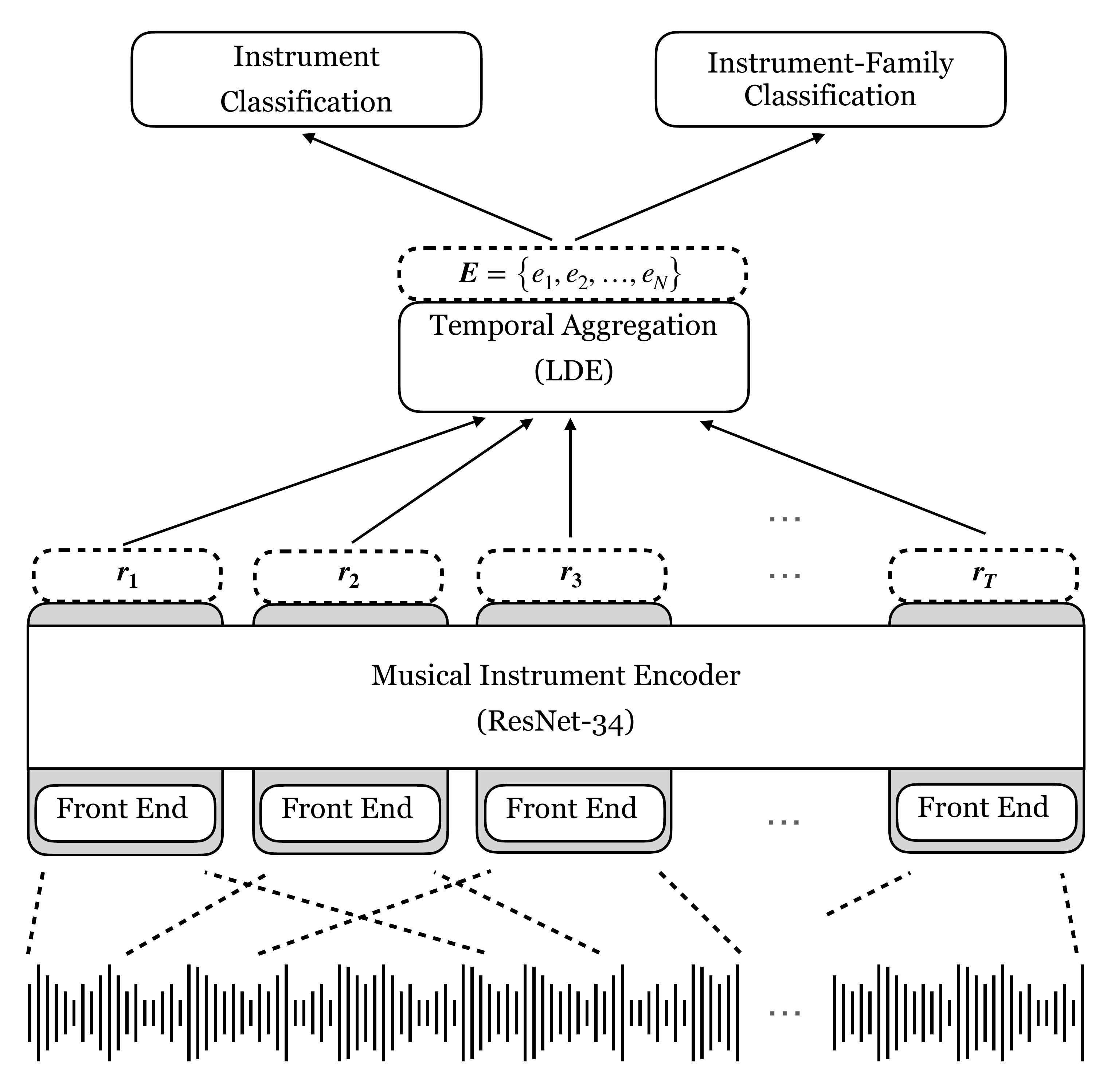} 
\caption{Architecture of proposed musical instrument recognition model inspired by speaker recognition. } 
\label{fig: arch} 
\end{figure}

\section{Musical Instrument Recognition Model Inspired by Speaker Recognition}
\label{sec:model}

As described earlier, we utilize popular speaker recognition technologies and their combinations for bench-marking performance on extracting representations of unseen instruments. Since a monophonic sound database is used, their system designs are similar to the original ASV ones, but we need to properly consider differences between speech and music signals. Here we describe how we amend the ASV technologies to cope with these differences. The overall architecture of the system is shown in Figure \ref{fig: arch}.

\subsection{Front-end}

The front-end transforms input audio into frame-level acoustic features. This may be done using spectral features like Mel-spectrogram and constant-Q transform (CQT) coefficients. Alternatively, learnable front-ends such as SincNet, RawNet \cite{jung2019rawnet}, and LEAF \cite{zeghidour21leaf} may be used to extract features from a waveform directly through neural networks. Here we describe how we amend SincNet, a popular ASV front-end, for extracting better features from the music signal. 

The front-end based on SincNet is implemented with a one-dimensional convolutional network whose kernel of each channel is regarded as a bandpass finite impulse response (FIR) filter. In the frequency domain, a bandpass filter for frequency $f$ is generally defined as the difference of two low-pass filters:
\begin{equation}
    G\left(f, f_{1}, f_{2}\right) = \mathrm{rect} \left(\frac{f}{2 f_{2}}\right)-\mathrm{rect}\left(\frac{f}{2 f_{1}}\right),
\end{equation}
where $\mathrm{rect(\cdot)}$ denotes a rectangular function in the frequency domain. We transfer a filter function into the time domain with an inverse Fourier transform to accommodate the waveform input. The filter function for a waveform sample $n$ in the time domain is given by:
\begin{equation}
    g\left(n, f_{1}, f_{2}\right)=2 f_{2} \mathrm{sinc}\left(2 \pi f_{2} n\right)-2 f_{1} \mathrm{sinc}\left(2 \pi f_{1} n\right),
\end{equation}
in which the $\mathrm{sinc}$ function is defined as $\mathrm{sinc}(x) = \sin(x) / x$. As can be seen from the equations, there are only two trainable parameters per filter band, $f_1$ and $f_2$, indicating the cutoff frequencies for the start and the end, and thus, the parameter size of the front-end can be significantly smaller than other fully-trained convolutional feature extractors \cite{Han17Deep}.

In the original SincNet paper, the initial cutoff frequencies $f_1$ and $f_2$ were set to those of the Mel filterbank \cite{Ravanelli18SincNet}. To handle music signals better, we adopt the concept of a scale used in the CQT transform and initialize learnable band-pass filters based on the musical scale. For instance, when we allocate one band-pass filter for each of the 128 MIDI note frequencies \cite{cooper21midi}, 
the corresponding $f_1$ and $f_2$ of the $k$-th filter are its two neighboring MIDI frequencies, which means $f_1 = 2^{((k-1)-69)/12}\times440$ and $f_2 = 2^{((k+1)-69)/12}\times440$, respectively. Those frequencies are further updated through back-propagation during the training phase. 

\subsection{Encoding and Temporal Aggregation}

The next step is to extract high-level features by down-sampling and transforming the output of the front-end and aggregating the frame-level features into a fixed-length embedding. Time delay neural network \cite{7846260}, Residual Network (ResNet) \cite{He16ResNet}, and Squeeze and excitation network \cite{hu2018squeeze} are frequently chosen for encoding, and statistical temporal pooling \cite{7846260}, attentive temporal pooling \cite{okabe18_interspeech} and  LDE pooling \cite{Cai18LDE} are dominant approaches for temporal aggregation. In our benchmark experiment, we employ ResNet34, a type of ResNet with 34 layers that is commonly used in speaker verification \cite{Li17deepspeaker}, for encoding. Since this ResNet34 module is a standard architecture, we briefly overview the temporal aggregation process only here. 

Since ResNet outputs a sequence of frame-level representations, it is important to transfer the sequential representation  $\boldsymbol{R} = \{\boldsymbol{r}_1, \boldsymbol{r}_2, ..., \boldsymbol{r}_T\}$ of length $T$ into a global time-invariant representation $\boldsymbol{E} = \{e_1, e_2, ..., e_N\}$ where $N$ is a fixed number regardless of the length of the input audio signals. This vector $\boldsymbol{E}$ can be used as an instrument embedding vector in multi-instrument audio synthesis. There are several popular techniques for temporal aggregation. Statistical temporal pooling \cite{7846260} conducts this process by computing the mean and standard deviation of the sequential representations. Attentive temporal pooling \cite{okabe18_interspeech} additionally adopts self-attention to select important segments before computing the statistics. 

In this paper, we employ the LDE method \cite{Cai18LDE}. Unlike temporal pooling, which implicitly assumes a uni-modal distribution, LDE adopts a clustering process and computes posterior probabilities of a fixed number of orderless learnable clusters via softmax. Each element of the fixed length $\boldsymbol{E}$ vector is represented as a product of the posterior and averaged residual vector of each cluster.  The collection of the learned cluster centers is a ``dictionary,'' and each learned cluster center is expected to represent characteristics of different regions in the sequence such as onset and overshoot regions, which is expected to help musical instrument recognition \cite{kai2019modeling}. 

\subsection{Objective Function and Output Layers}

The fixed-length vector $\boldsymbol{E}$  is used to predict the instrument category included in the training database based on the softmax, and backpropgation is conducted to update all  LDE, ResNet34 and SincNet parameters. However, this does not guarantee that different instruments in the same instrument family have relatively similar embedding vectors since the network may try to predict the instrument types without considering their instrument family. Therefore, as Figure \ref{fig: arch} shows, we also introduce instrument-family prediction as an additional regularization task to constrain the embedding space and make it more intuitive.  

Moreover, we adopt angular softmax (A-softmax), a discriminative variant of softmax that explicitly considers and enlarges the angular margin between classes. This has obtained high performance recently in face recognition and speaker recognition tasks \cite{Huang18ASoftmax, Xiang19margin}.  

\section{ASV benchmarking techniques for verification of unseen musical instruments}
\label{sec:exp}

\subsection{Data}
\label{data_overview}

The NSynth dataset \cite{nsynth2017} and the RWC musical instrument dataset \cite{Goto03RWC}, two well-known databases in music processing, are used in this paper. An overview of the datasets is shown in Table \ref{tab:data}.

In the NSynth Database \cite{nsynth2017}, individual notes are played by various instruments at different pitches and velocities, which makes it an appropriate choice for constructing a comprehensive latent space of musical instrument sounds.
The NSynth dataset provides pre-defined training, validation, and evaluation database partitions. 
Instrument-family categories in the validation and testing sets all exist in the training set, while individual instruments in the training set have no overlap with those in the validation and testing sets. Thus, from the perspective of instrument-family classification, the categories in validation and testing sets are \textit{all seen}; however, from the perspective of instrument classification, the categories in the validation and testing sets are \textit{all unseen}, and hence the task of predicting these instruments is not identification but verification. 

As for the RWC database, each sample of a musical instrument includes a continuous performance of all notes in the pitch range of the instrument. Since there are no pre-defined database partitions, we adopted a similar partition method to NSynth. We selected a number of different variations of musical instruments 
to form validation and testing sets such that the instrument family categories in the validation and testing sets are all seen, but instrumental variations in the validation and testing sets are all unseen.  Instrument variations are different combinations of instrument manufacturer and performer.

\begin{table}[tb]
    \centering
\caption{Overview of two musical instrument databases used.}
\begin{tabular}{lcc}
\toprule
& \textbf{NSynth} & \textbf{RWC instrument} \\
\midrule
Categories in database & 1,006 & 45 \\
Categories in training set & \phantom{0}953 & 45 \\
Categories in valid.\ set & \phantom{00}53 & 30 \\
Categories in test set & \phantom{00}53 & 30 \\
\midrule
Samples in database & 305,979 & 1556 \\
Samples in training set & 289,205 & 1144 \\
Samples in valid.\ set & \phantom{0}12,678 & \phantom{0}209 \\
Samples in test set & \phantom{00}4,096 & \phantom{0}203 \\
\midrule
Total playback time (with silence) & 340.0 hours & 71.7 hours \\
\bottomrule
\end{tabular}
\label{tab:data}
\end{table}

\subsection{Experimental Conditions}

We first study how well the proposed instrument recognition model performs. More specifically, we show the verification results of the unseen instruments and instrument variations included in test sets of the NSynth and RWC datasets and analyze whether the ASV techniques described earlier can improve verification of the unseen instruments or not.

\subsubsection{DNN Training}
All systems in our experiments relied on ResNet34. The input to the system was 3- to 5-second segments of musical instrument performances. We initialized the front-end with either the Mel filterbank or the CQT filterbank. The number of filters for the CQT filterbank was set to 122 since one filter was allocated per MIDI note. This filterbank can cover at least the second harmonics of the highest pitch ($F\#7$) of the RWC database, and the cut-off frequency of the highest filterbank  is close to the Nyquist frequency for the NSynth database. For the Mel filterbank, we consider both the case of using the same number of filters as the CQT one and 80 filters, which is the standard setting frequently used in speaker recognition. In the training stage, the low frequency $f_1$ and the frequency band $|f_1 - f_2|$ are the two trainable parameters of each channel in the front-end. 
Considering the covered frequency range and frequency resolution, the minimum values of the low frequency and frequency band of each channel were set to 5 Hz and 5 Hz, respectively, so that the generated representation was interpretable in terms of frequency. 
For the LDE, we set the number of dictionary clusters $C = 32$ and the dimension of the embedding vector $\boldsymbol{E}$ is $N=512$. The angular margin used for A-softmax is set to 2.

We trained the models with the Adam optimizer for 30 epochs to reach a point of convergence. Specifically, the models were trained with a scheduler to dynamically adjust the value of the learning rate -- the learning rate linearly increased in the first 8000 steps followed by an exponential decrease.  We run training for each model three times with the random seed $10^{k+1}$ on the $k-$th run, as the churn resulting from initialization, data preprocessing, and other random processes  affects the consistency of models, which has been demonstrated in several recent works of literature from various fields \cite{wang2021comparative, madhyastha2019model, bhojanapalli2021reproducibility}. Multiple runs with assigned seeds ensures reproducibility and increases the stability of the experiments.  Experimental results of each condition are presented by the mean and standard deviation values of the three runs in Tables \ref{tab:ablation} and \ref{tab:front-end}.   On the NSynth dataset, all systems were trained from scratch with different random seed values three times\footnote{Source code for our experiments is available at \url{https://github.com/Alexuan/musical_instrument_embedding}}. Since the RWC dataset is smaller than the NSynth dataset, all three models trained on the NSynth dataset with different random seed values were fine-tuned on the RWC dataset. Due to the difference in categories between the two datasets, we replaced the dual output layer of the pre-trained models with randomly-initialized new output weights. During the fine-tuning phase, all model parameters including the front end, encoder network, and pooling and output layers were updated. 

\subsubsection{Data Augmentation}
In many recognition tasks including speaker recognition, data augmentation is often useful for improving the robustness of predictive models. Augmentation is done by adding varying degrees and types of noise, increasing or slowing down the speed of the audio, or other modifications. However, for the task in this paper, the aforementioned methods are likely to change the characteristics of a musical instrument sound and provide incorrect information for model optimization. We therefore adopted a straightforward yet efficacious method to create new samples by trimming the silences of samples\footnote{In the NSynth dataset, since the individual note sounds of many instruments do not last for the entire four seconds, there were a large number of silent segments on the ends of the input samples.} in the NSynth dataset and then randomly concatenating samples from the same musical instrument. Both individual and concatenated continuous notes were equally mixed during training in order to eliminate the domain gap between them.

\subsubsection{EER calculation}
For computing the EERs\footnote{We computed EERs with pyBOSARIS available at \url{https://gitlab.eurecom.fr/nautsch/pybosaris}}, we first chose five audio samples per instrument as enrollment data and computed the averaged instrument embedding vector per instrument. Then, from a pool of randomly selected samples with 20 samples from the target instrument and 20 samples from non-target instrument categories, we chose one sample, computed its embedding vector, calculated its cosine distance to the average enrollment embedding vector of the target instrument, repeated this process for all samples in the pool, and computed scores required for computing EERs.

\subsection{Experimental Results}
\label{exp1}

To evaluate the generalization performance and the robustness of the ASV techniques described earlier, we conducted all experiments on both the NSynth and the RWC databases which contain different sets of unseen musical instruments or instrumental variations as the test sets. Since the number of all combinations of experimental settings is too large, we divided comparisons into two parts. In the first part, we focus on the training manner of the neural network and compare and evaluate the impacts of data augmentation, regularization based on instrument family prediction and A-softmax using the same network architecture and same features. In the second part, we focus on front-end factors such as frequency scales and number of filters and compare different features using the same network architecture and same training manner. 

\begin{table}[t]
\caption{\it Ablation study on the training configurations.  We excluded one of the components (data augmentation, regularization based on instrument family prediction and A-softmax) from a base system that uses data augmentation, SincNet initialized based on CQT filters, LDE and ResNet34.}
\label{tab:ablation}
\centering
\begin{tabular}{l|c|c|c}
\toprule
        & NSynth   & RWC   & Average \\ 
System  & EER (\%) $\downarrow$ & EER (\%) $\downarrow$ & EER (\%) $\downarrow$ \\          
\midrule 
Base                        & 3.74 $\pm$ 0.37 & 1.03 $\pm$ 0.11 & 2.39 \\
Base w/o data augmentation  & 7.58 $\pm$ 2.34 & 2.22 $\pm$ 0.29 & 4.90 \\
Base w/o instrument family  & 3.14 $\pm$ 0.34 & 1.12 $\pm$ 0.20 & 2.13 \\
Base w/o A-softmax          & 4.60 $\pm$ 0.23 & 2.44 $\pm$ 0.08 & 3.52 \\
\bottomrule
\end{tabular}
\end{table}
\begin{figure}[t]
\centering
\includegraphics[width=0.68\columnwidth]{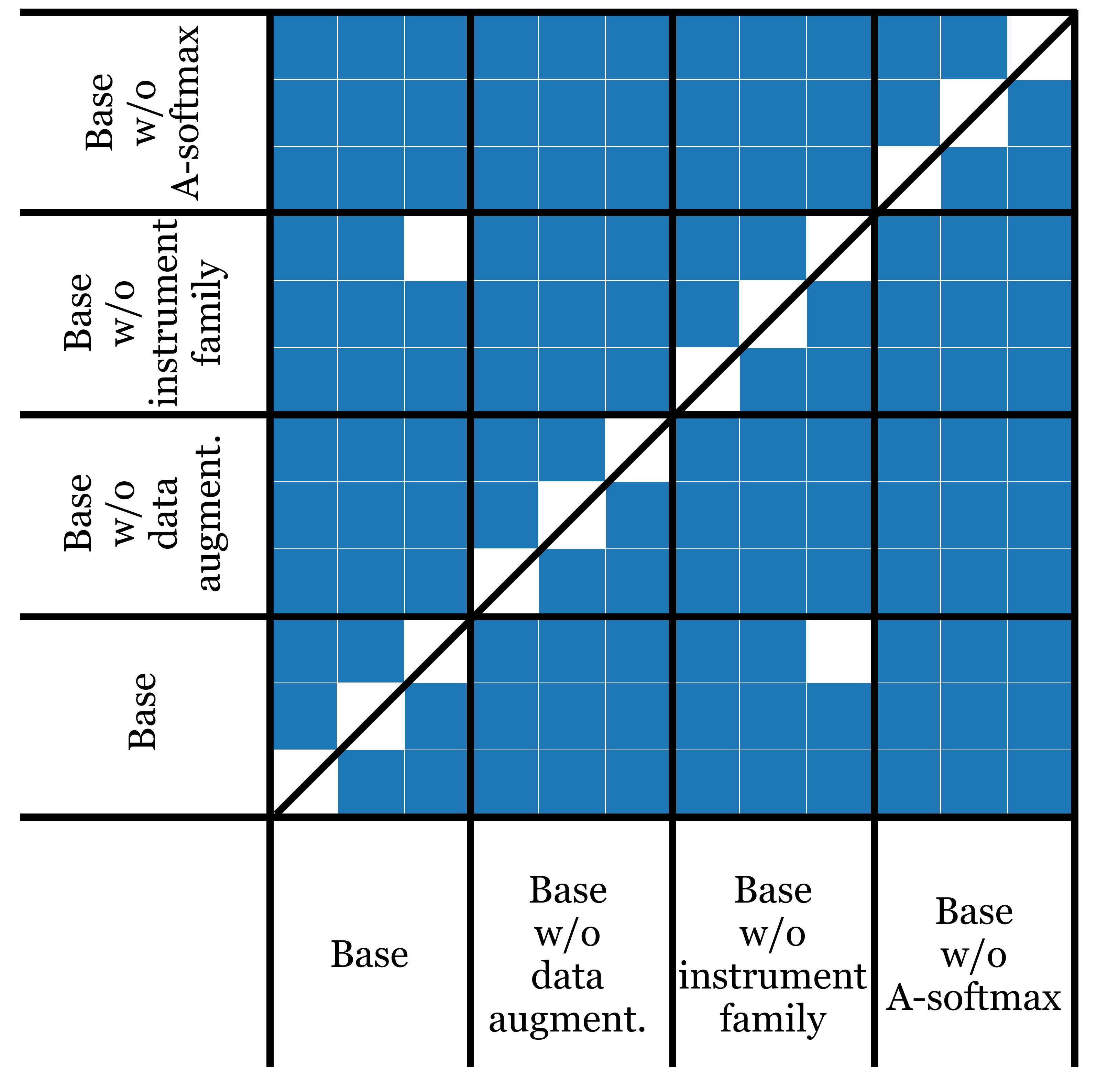}\\
(a) Statistical significance test results on the NSynth dataset \\
\vspace{5mm}
\includegraphics[width=0.68\columnwidth]{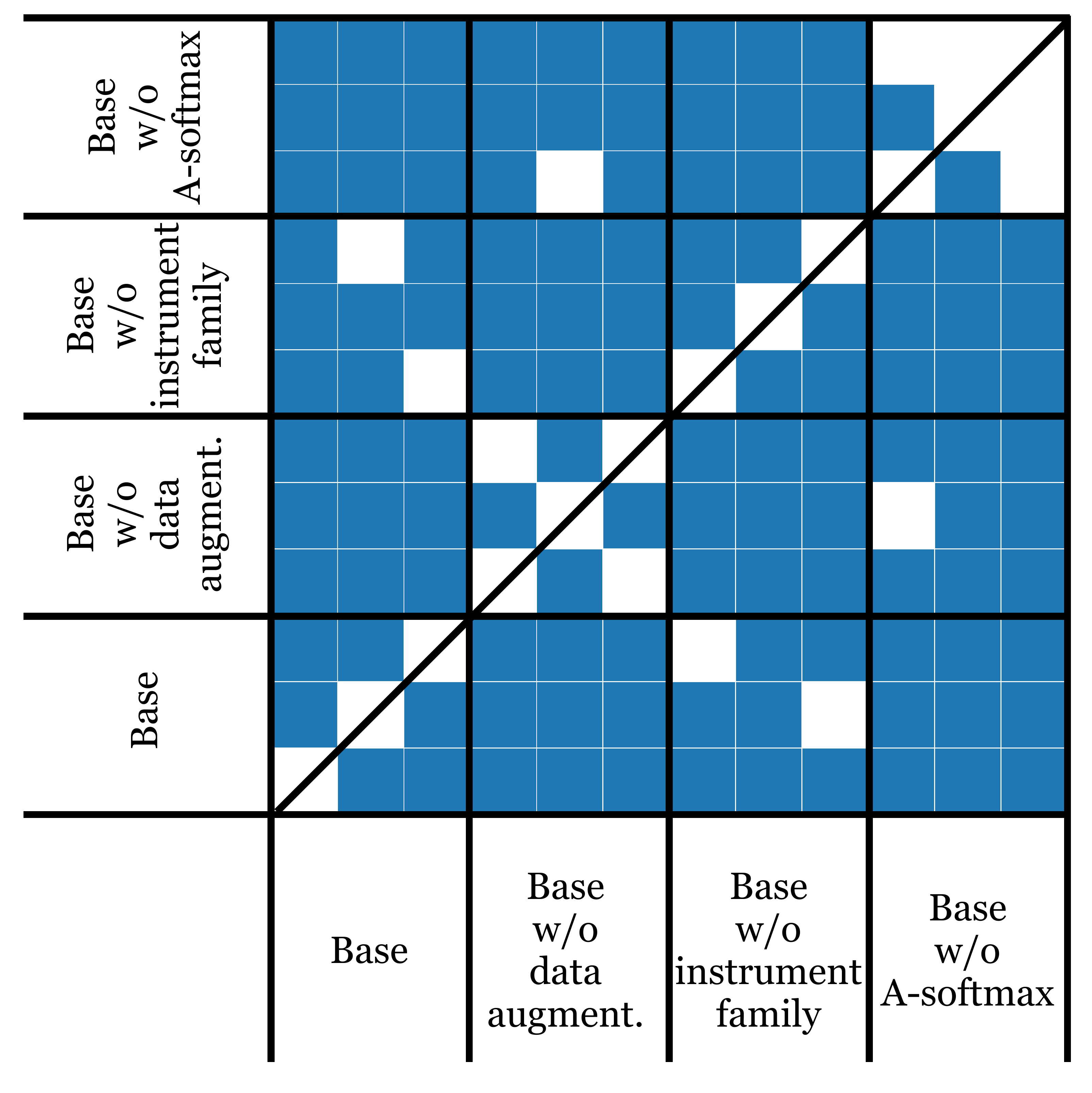} \\
(b) Statistical significance test results on the RWC dataset \\
\caption{Statistical significance test on EERs for training strategy experiments given $\alpha = 0.05$ with Holm-Bonferroni correction. Blue shade represents significant difference, while white data points indicate $(A,B)$ models were not statistically different. Four systems were tested on the NSynth Database for statistical significance, with each system having three entries with three different random seeds. }
\label{fig:statistical_significance_eer2}
\end{figure}

Table \ref{tab:ablation} shows the first ablation study on the training manner of the neural network. We excluded one of the components (data augmentation, regularization based on instrument family prediction and A-softmax) from a base system that uses data augmentation, SincNet initialized based on CQT filters, LDE and ResNet34 and trained each variant three times with different random seeds. 
Figure \ref{fig:statistical_significance_eer2} shows statistical significance test results on EERs. Since we have three models trained with different random seeds for each condition, we analyzed both significant differences for models within each condition and significant differences between models in different conditions.

From the table, we first see that the base system has reasonable performance. Its EERs are only 3.74\% and 1.03\% on the NSynth and RWC datasets, respectively. But, as expected, we also see that its performance depends on the weight initialization and its random seed values used since EER differences for three models within the base condition are all statistically significant ($p<0.0001$) on both the NSynth and RWC datasets. Next, we see that data augmentation and A-softmax have an important role since averaged EERs become worse -- from 2.39\% to 4.90\% and to 3.52\%, respectively. Their EER degradations are all statistically significant ($p<0.0001$) for any model pairs of the base condition and base without data augmentation, and also any model pairs of the base condition and base without A-softmax, on both the NSynth and RWC datasets.

Further, we also see that the condition without regularization based on instrument family prediction has slightly better EER results than the base condition. They are statistically better for 8 out of 9 model pairs on the NSynth dataset and 2 out of 9 model pairs on the RWC dataset. Since the instrument family prediction acts as a kind of regularization, this is a  natural consequence, but, on the other hand, it is expected that the embedding space becomes less intuitive in terms of the instrument family category. This will be investigated further in the next section. 

\begin{table}[t!]
\caption{\it Verification results of unseen instruments with different front-end features. All systems use data augmentation, ResNet34, LDE and A-softmax.}
\label{tab:front-end}
\centering
\begin{tabular}{l|c|r|c|c|c}
\toprule
\multicolumn{3}{c|}{Filter} & NSynth     & RWC                   & Average  \\ 
type       & update  & number  & EER (\%) $\downarrow$ & EER (\%) $\downarrow$ & EER (\%) $\downarrow$ \\          
\midrule 
Mel  & & 80                & 3.19 $\pm$ 1.01 & 1.89 $\pm$ 0.23 & 2.54 \\
Mel  & $\checkmark$ & 80   & 3.42 $\pm$ 0.18 & 2.26 $\pm$ 0.15 & 2.84 \\
Mel  & & 122               & 3.44 $\pm$ 0.50 & 2.18 $\pm$ 0.18 & 2.81 \\
Mel  & $\checkmark$ & 122  & 3.54 $\pm$ 0.14 & 1.98 $\pm$ 0.11 & 2.76 \\
CQT  &              & 122  & 4.19 $\pm$ 0.30 & 0.96 $\pm$ 0.13 & 2.58\\
CQT  & $\checkmark$ & 122  & 3.74 $\pm$ 0.37 & 1.03 $\pm$ 0.11 & 2.39 \\
\bottomrule
\end{tabular}
\end{table}

\begin{figure}
\begin{center}
\includegraphics[width=0.68\columnwidth]{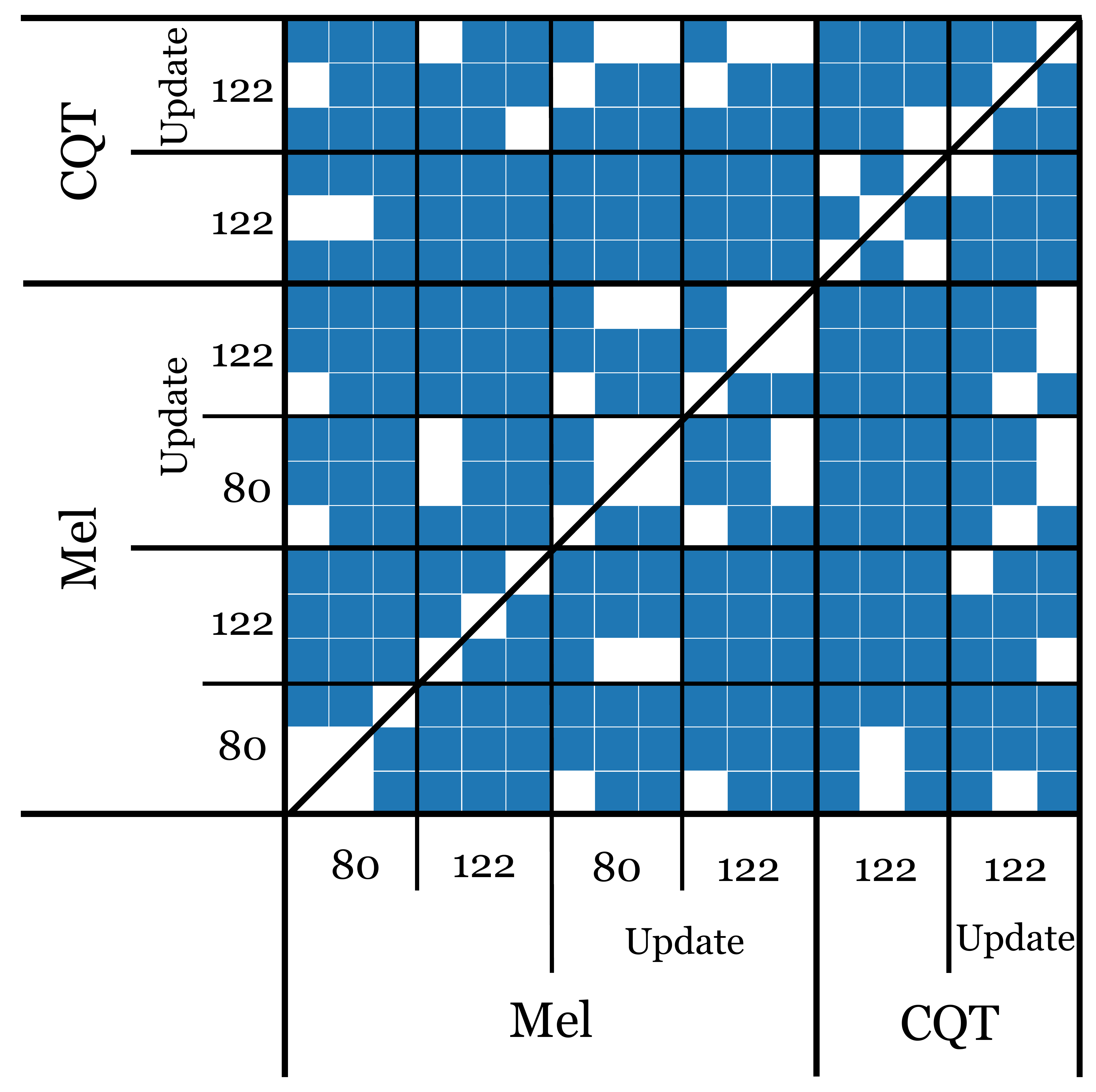} \\
(a) Statistical significance test results on the NSynth dataset \\
\vspace{5mm}
\includegraphics[width=0.68\columnwidth]{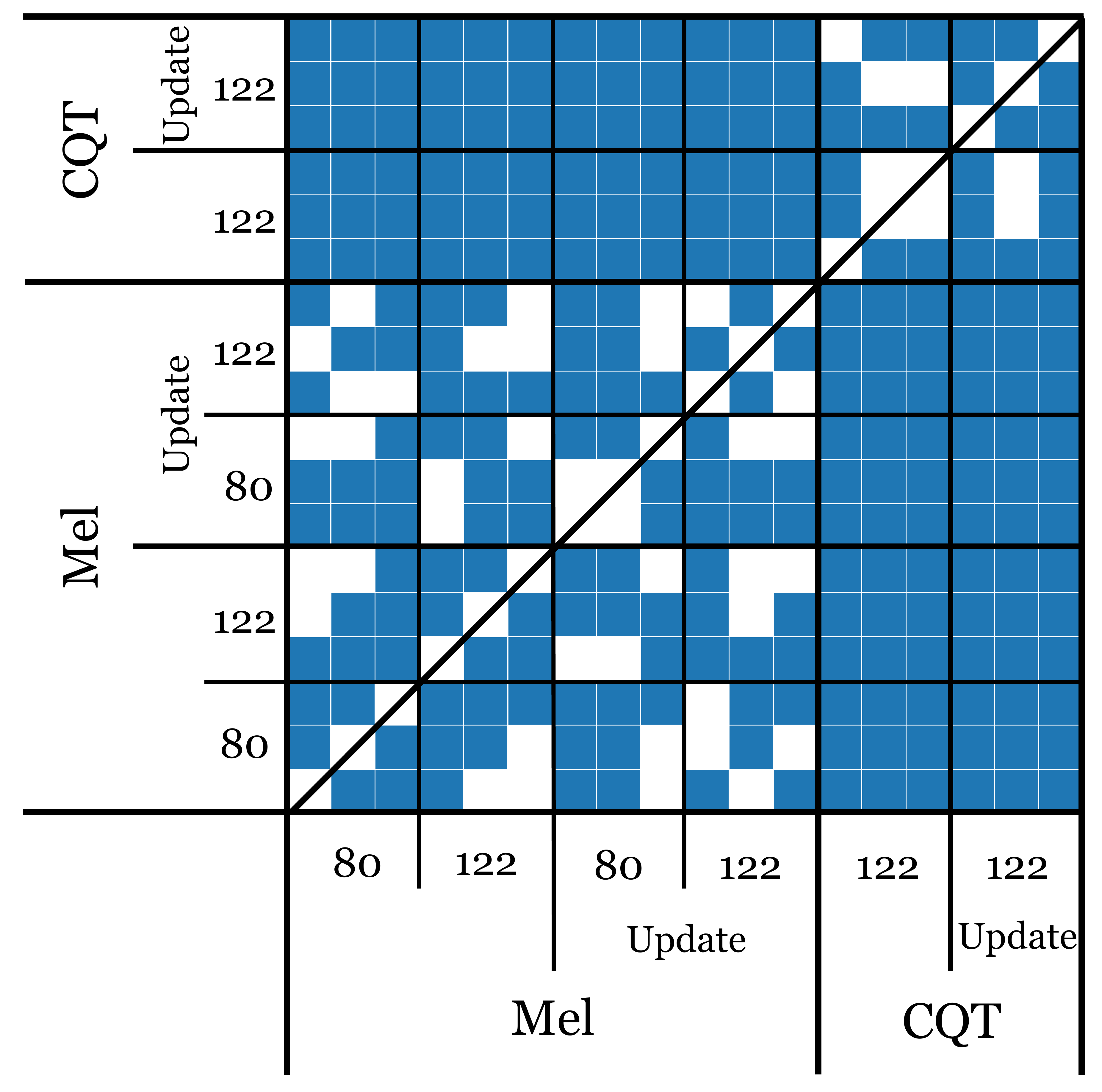}\\
(b) Statistical significance test results on the RWC dataset \\
\end{center}
\caption{Statistical significance test on EERs for front-end feature experiments given $\alpha = 0.05$ with Holm-Bonferroni correction. Blue shade represents significance difference, while white data points indicate $(A,B)$ models were not statistically different. Six systems were tested on the NSynth Database for statistical significance, with each system having three entries with three different random seeds. }
\label{fig: statistical_significance_eer}
\end{figure}

Table \ref{tab:front-end} shows verification results using different front-end features. All systems use the same ResNet34 network with data augmentation, LDE and A-softmax. We analyzed three factors related to SincNet based front-end feature extraction and they are a) initial frequency scales used for the filterbanks, b) updating of cutoff frequencies of the filterbanks, and c) number of filters.  

We see that a system that uses SincNet initialized based on CQT filters is good on average. But there is no strong clear pattern on the NSynth dataset and this is also underpinned by the statistical significance test results in Figure \ref{fig: statistical_significance_eer} (a) in which there are many model pairs where differences are not statistically significant on the NSynth dataset. But, interestingly, we can see that all systems using the CQT filters work well on the RWC dataset, and all pairs of CQT-based models and Mel-based models have statistically-significant differences ($p<0.0001$) as we can see from Figure \ref{fig: statistical_significance_eer} (b). This interesting tendency will be investigated further in the next section.

\section{Complementary Evaluation through Analysis of Encoded Information}
\label{exp3}

\subsection{Motivation}

\begin{figure}[t]
\begin{center}
	\includegraphics[width=1.0\columnwidth]{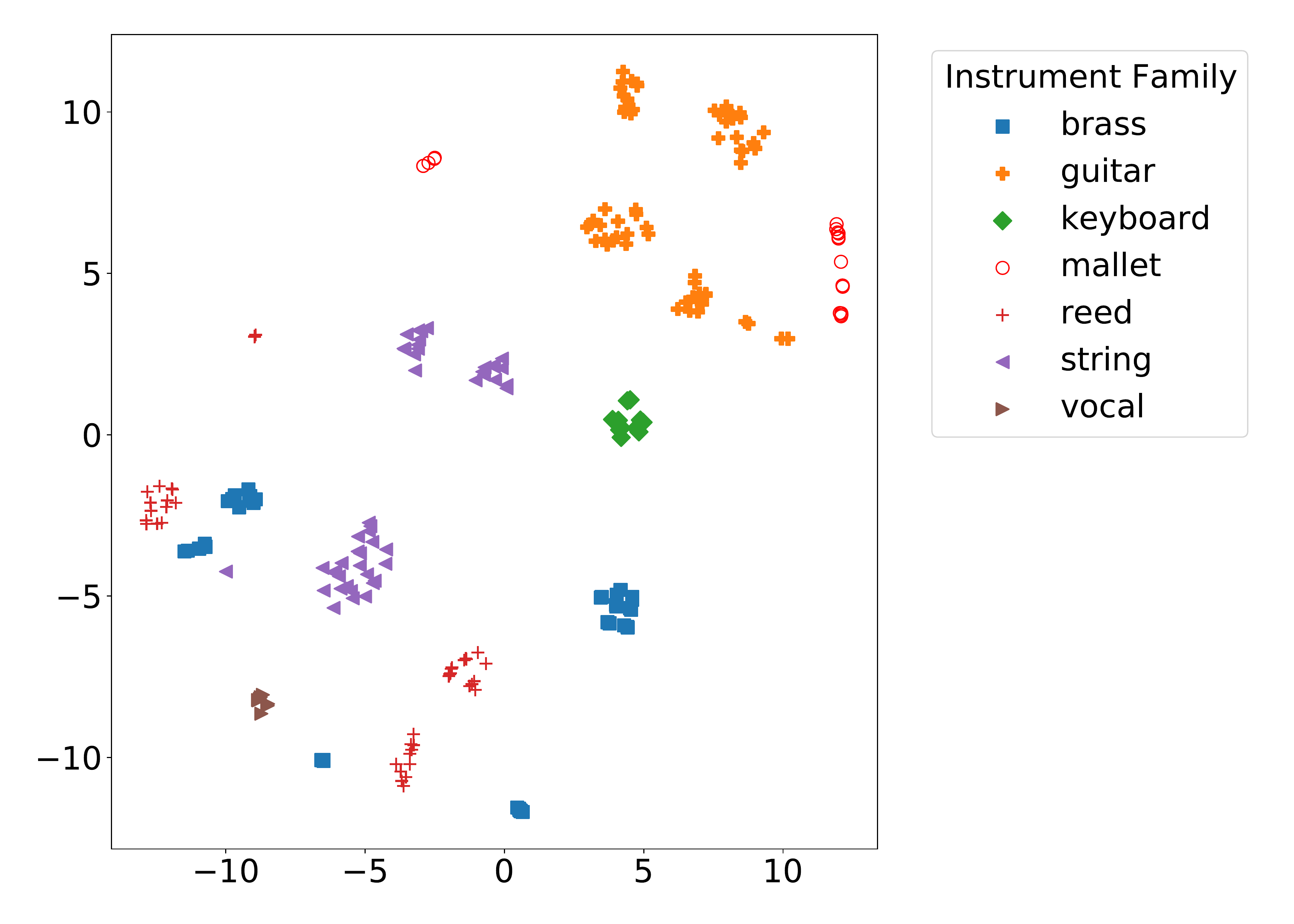} \\
(a) Embeddings obtained from the instrument encoder trained without using the instrument-family category prediction\\
\vspace{2mm}
\vspace{2mm}
	\includegraphics[width=1.0\columnwidth]{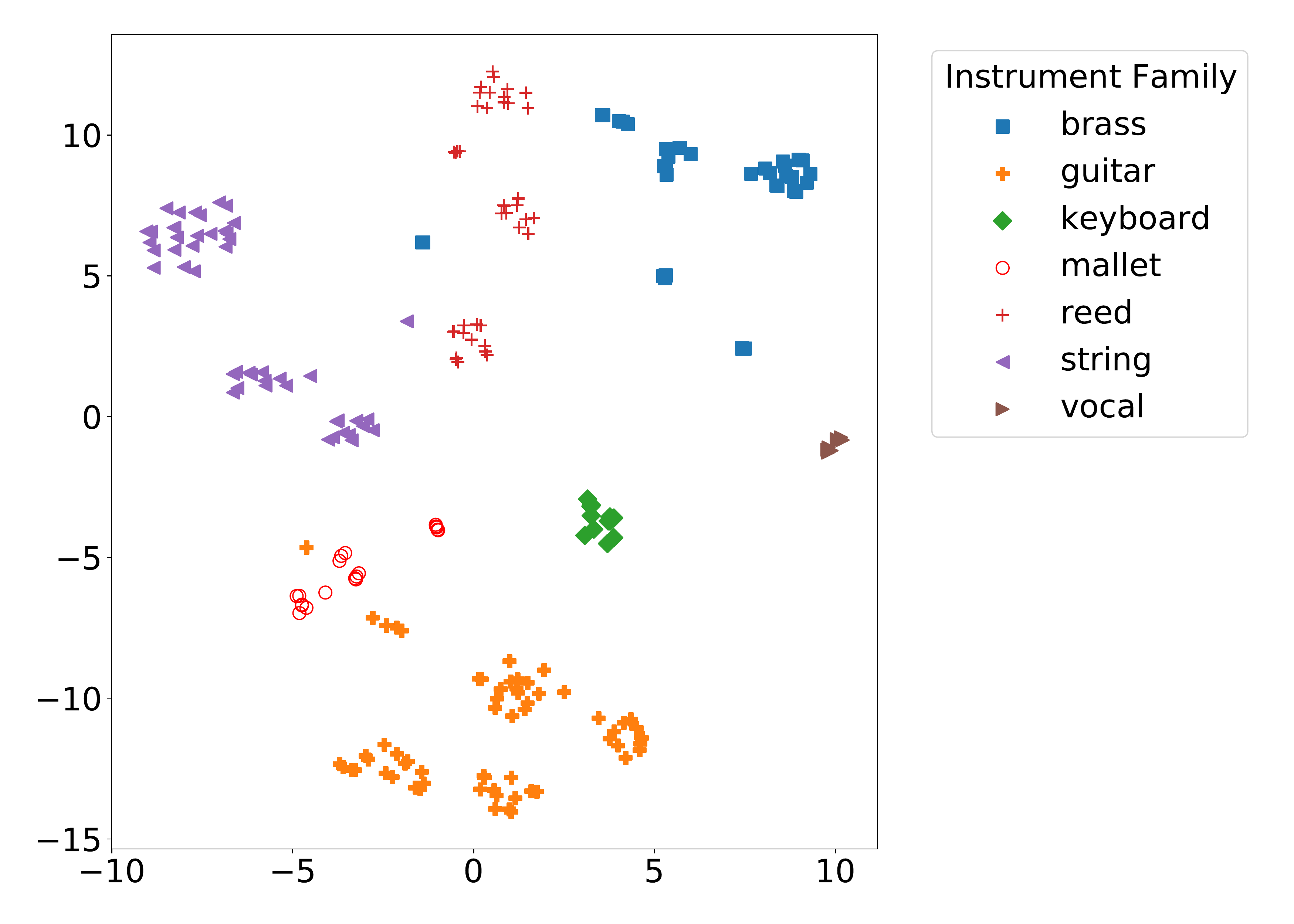} \\
(b) Embeddings obtained from the instrument encoder trained using the instrument-family category prediction as an additional regularization term\\
\end{center}
\caption{t-SNE based visualization of embedding vectors of multiple unseen NSynth instruments having the same pitch.}
\label{fig: visualization}
\end{figure}

Through the experiment in the previous section, we demonstrated that EER-based evaluation is useful for finding a suitable architecture and training techniques for acquiring appropriate embeddings of unseen instruments. However, it is not possible to evaluate all perspectives with EER alone. For instance, it is expected that adding instrument-family prediction to the objective function has the effect of bringing the embedding vectors of instruments in the same category closer together, but this cannot be measured via EER and hence complementary assessments need to be conducted.

Figure \ref{fig: visualization} shows t-SNE based visualizations of embedding vectors obtained from the instrument encoders trained with and without the additional instrument-family regularization term. Each point represents a sample from a different instrument having the same pitch value. From Figure \ref{fig: visualization} (b), we can visually confirm that, for instance, the embedding vectors of instruments in the string, reed, and brass families are all located closer together compared to its counterpart (a). 

In addition to visual inspection, building shallow classifiers using the embedding vectors is also expected to provide us with complementary insights into the vectors objectively. For example, shallow classifiers using embedding vectors derived from the instrument encoder trained using instrument-family category prediction as the regularization would be expected to have better prediction performance of the instrument-family label.\footnote{In fact, we ran the instrument-family probing task on embeddings extracted from the model trained with and without instrument-family regularization. We observed that a decision tree classifier is able to achieve $7.4\%$ relative improvement at this task on NSynth and $5.5\%$ on the RWC Database by using the instrument-family regularized embeddings, indicating that instrument family information is made more prominent and therefore accessible to the weaker classifiers by using the multi-task learning.}

Thus, we devised a series of experiments to probe the encoded information with regard to the available metadata. Inspired by \cite{Raj19probing} presented by Raj \MakeLowercase{\textit{et al.}}, we trained simple shallow classifiers using the embeddings to predict the different types of metadata labels included in the NSynth and RWC datasets, respectively, on the basis of the premise that if information about these labels is present in the embeddings, then learning a classifier should be possible. The shallow classifiers used in these probing tasks were multilayer perceptron (MLP), Support Vector Machine (SVM), and Decision Tree (DT), and they were trained using a machine learning toolkit, scikit-learn\footnote{https://scikit-learn.org/stable/index.html}. Since the amount of training data used for the shallow classifiers was small, and since the relationship between the embeddings and metadata was not clear, we investigated the classification performance over different types of classifiers. Note that we used two databases for our experiments, but not all types of metadata existed in both databases. Therefore, some of the probing tasks were implemented using only one of the databases.

\begin{figure*}[h!]
\centering
\begin{subfigure}[b]{0.45\linewidth}
\centering
\includegraphics[width=\linewidth]{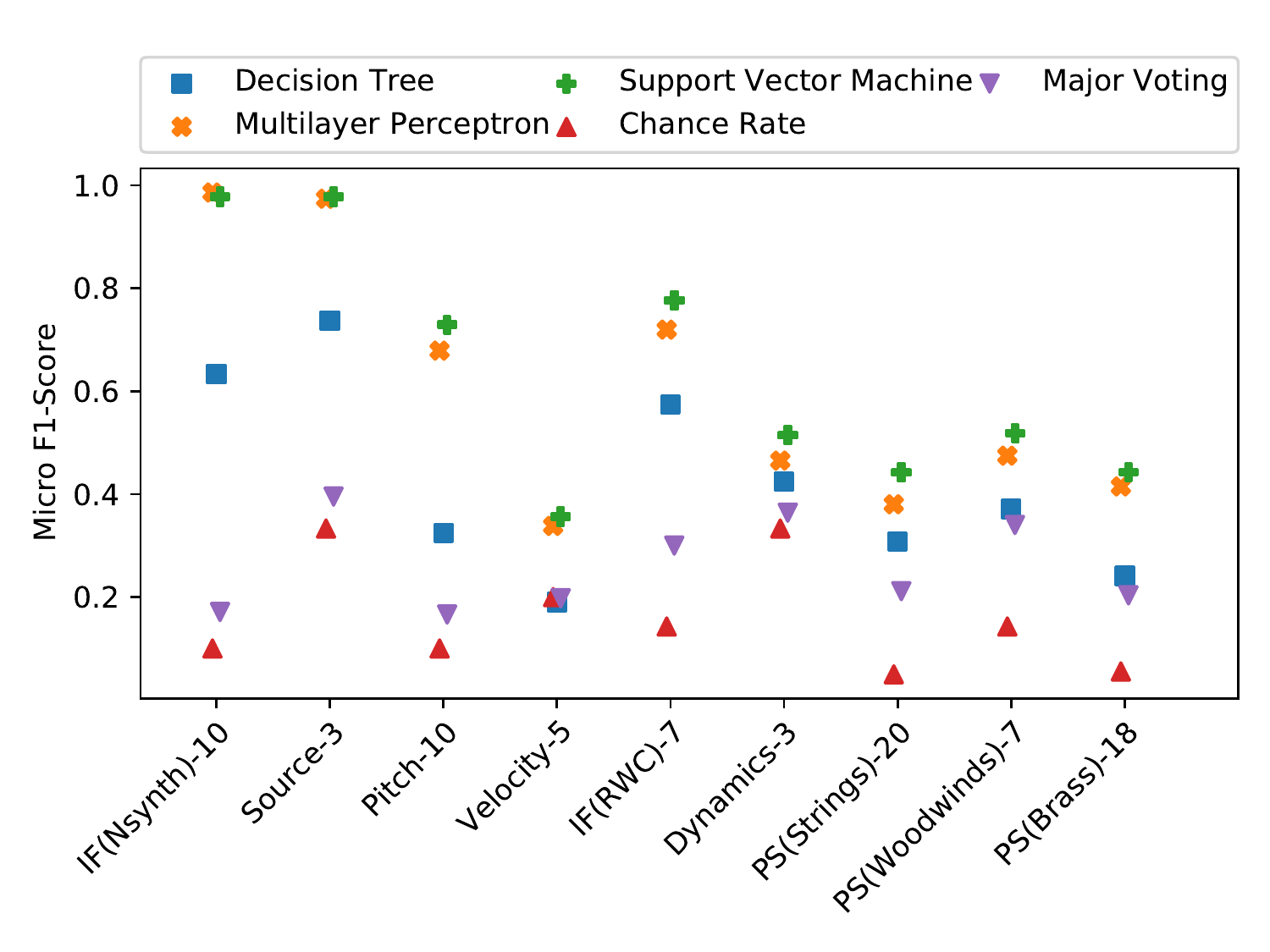}
\caption{Initialization using the Mel filterbank}
\label{fig: probing_mel}
\end{subfigure}
\begin{subfigure}[b]{0.45\linewidth}
\centering
\includegraphics[width=\linewidth]{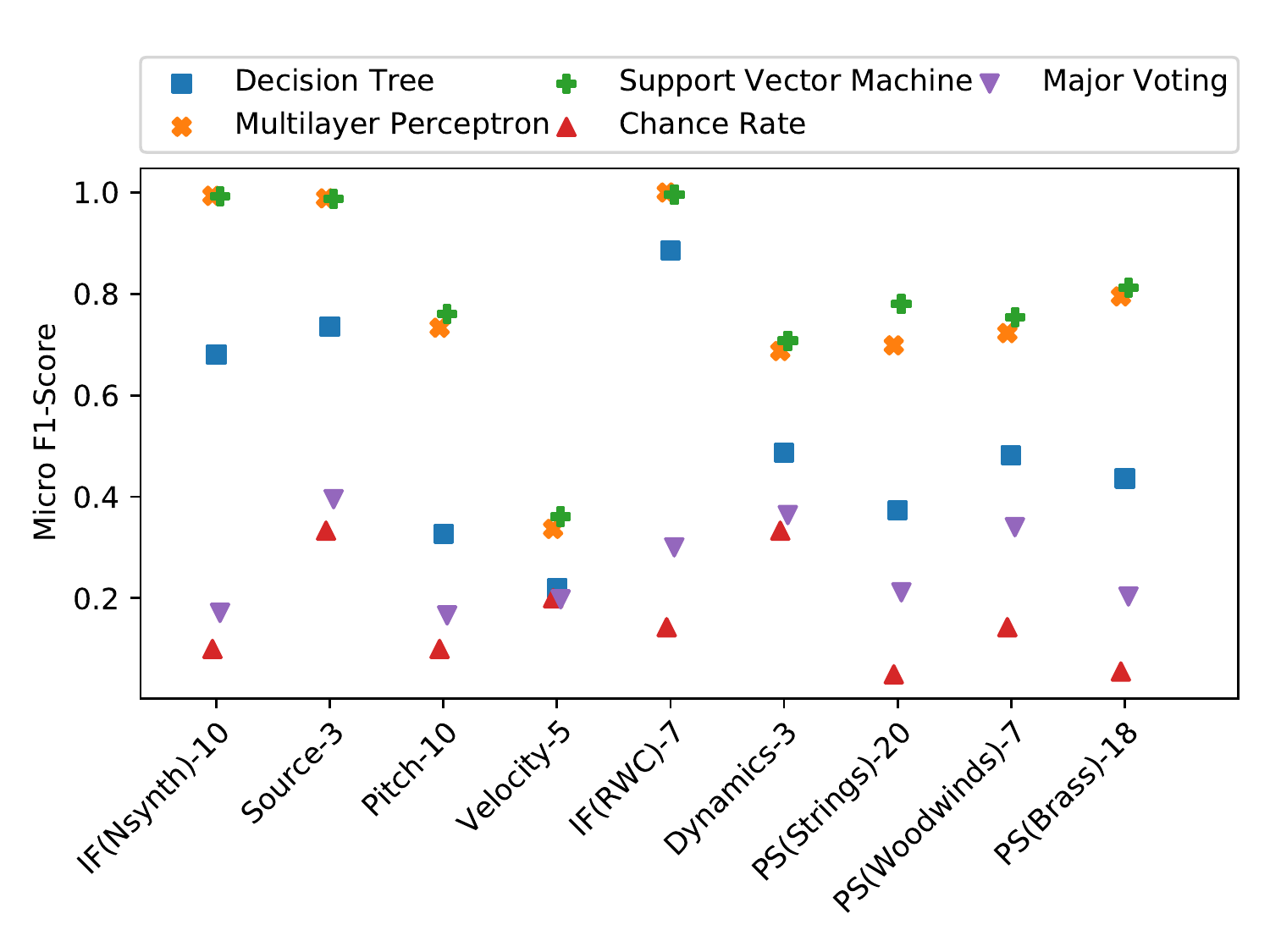}
\caption{Initialization using the CQT filtebank}
\label{fig: probing_midi}
\end{subfigure}
\caption{Prediction results using embeddings extracted from two systems using different SincNet initializations. ``IF" and ``PS" are abbreviations for ``Instrument Family" and ``Playing Style", respectively.}
\label{fig: probing}
\end{figure*}

\subsection{Details of Shallow Classifiers}
For the shallow classifiers, we adjusted their configurations as follows. For MLPs, considering the relationship between degree of freedom and number of data points, we set the hidden size of MLP to 300 and 100 for the NSynth and RWC datasets, respectively. For the SVM models, we used a radial basis function kernel and searched for a proper regularization parameter $C$ by model selection. The search range of $C$ was set to $[10^{-3}, 10^{3}]$ with a log scale. Taking into account that there are many hyper-parameters for DTs, such as maximum depth, minimum number of leaf nodes, and criterion, we leveraged grid search to select a group of proper hyper-parameters. For other adjustable parameters of MLP, SVM, and DT, we used default configurations of scikit-learn. 

For training the shallow classifiers for probing tasks, it is necessary to have an additional training set separate from the one used for training the embedding model. To do this, the original test sets of the NSynth and RWC datasets were divided into three disjoint subsets with a ratio of 8:1:1, respectively, and they were used as the training, validation, and new test sets for the probing tasks.

\subsection{Metadata Labels Used for Probing Tasks}

We used the following discrete metadata labels available for the NSynth dataset and RWC dataset below in addition to the instrument-family category: 
\begin{enumerate}
    \item Source (3 classes)
    \item Pitch (10 classes)\footnote{To make comparisons with other metadata labels fair, we quantized them per octave.}
    \item Velocity (5 classes)
    \item Dynamics (3 classes)
    \item Playing Style (20, 7, and 18 classes)
\end{enumerate}

For instrument-family, source, pitch, velocity, and dynamics, instrument-independent classifiers were trained.  For playing style, we chose instruments from the same instrument family that usually share similar playing styles (e.g.,\ \MakeLowercase{\textit{Spiccato}}, \MakeLowercase{\textit{Tremolo}}, and \MakeLowercase{\textit{Pizzicato}} of the string family) and trained a classifier per instrument family. We trained playing-style classifiers for three instrument families: strings (bowed), woodwinds, and brass. There are 20 playing-style classes in strings (bowed), 7 classes in woodwinds, and 18 classes in brass, respectively. 
For more details of the metadata labels, please see \cite{nsynth2017,Goto03RWC}. 

\subsection{Experimental Results}

We can use instrument embedding vectors obtained from any instrument encoder for the probing tasks, but, in this paper we selected two variants using different SincNet initializations based on Mel or CQT filters. This is because these two variants show inconsistent results on the NSynth dataset and all systems using the CQT filters worked better on the RWC dataset as shown in Table \ref{tab:front-end} of Section \ref{sec:exp}, and we are interested in finding out the potential reasons. 

We evaluate classifier performance using Micro and Macro F1-scores.  The F1-score is the harmonic mean of precision and recall.  The Micro F1-score, also known as accuracy, is a basic measurement that calculates the ratio of correctly predicted samples to the total number of samples.  Both the NSynth and RWC datasets have a class imbalance, where the number of samples among different instrument families is quite different. We therefore computed the Macro F1-score as well, which sums the F1-Scores for each class without weighting them.  This penalizes poor performance on minority classes.

Figures \ref{fig: probing} (a) and (b) show the Micro F1-score results of the probing tasks using embeddings extracted from the two systems using different SincNet initializations. Since the numbers of output classes for the individual probing tasks were different from each other, we cannot compare the raw results across probing tasks directly. Therefore, Tables \ref{tab:probing-relative} (a) and (b) show the results for the relative improvements in the F1 scores of individual shallow classifiers compared with majority voting, that is $I = \frac{F1_{\mathrm{classifier}} - F1_{\mathrm{majority}}}{F1_{\mathrm{majority}}}$. 

\begin{table*}[t]
\centering
\caption{Relative improvements in F1 scores compared with majority voting. Relative improvement for each classifier was computed as $I = \frac{F1_{\mathrm{classifier}} - F1_{\mathrm{majority}}}{F1_{\mathrm{majority}}}$.} 
(a) Relative improvements using embeddings extracted from an encoder using SincNet initialization based on the Mel filterbank\\
\begin{tabular}{l|c|cc|cccccc}
\toprule
& & \multirow{2}*{NSynth} & \multirow{2}*{RWC} & \multicolumn{2}{c}{MLP} & \multicolumn{2}{c}{SVM} & \multicolumn{2}{c}{DT}\\
        \cline{5-10}
& \#classes & & & Micro F1 $\uparrow$ & Macro F1 $\uparrow$ & Micro F1 $\uparrow$ & Macro F1 $\uparrow$ & Micro F1 $\uparrow$ & Macro F1 $\uparrow$ \\
\midrule
Instrument Family (Nsynth) & 10 & $\surd$ & & 4.78 & 32.67 & 4.73 & 32.31 & 2.71 & 18.85  \\
Source & 3 & $\surd$ & & 1.46 & 4.13 & 1.48 & 4.18 & 0.87 & 2.87 \\
Pitch & 10 & $\surd$ & & 3.09 & 18.38 & 3.40 & 20.31 & 0.96 & 7.51 \\
Velocity & 5 & $\surd$ & & 0.71 & 4.14 & 0.80 & 4.52 & -0.04 & 1.70 \\
Instrument Family (RWC) & 7 & & $\surd$ & 1.16 & 8.01 & 1.33 & 9.10 & 0.72 & 4.65  \\
Dynamics & 3 & & $\surd$ & 0.30 & 1.67 & 0.42 & 1.92 & 0.17 & 1.36 \\
Playing Style (Strings) & 20 & & $\surd$ & 0.80 & 13.21 & 1.09 & 13.22 & 0.46 & 5.22 \\
Playing Style (Woodwinds) & 7 & & $\surd$ & 0.40 & 4.15 & 0.52 & 5.13 & 0.09 & 1.70 \\
Playing Style (Brass) & 18 & & $\surd$ & 1.04 & 14.7 & 1.18 & 13.55 & 0.19 & 5.14 \\
\bottomrule
\end{tabular} \\
\vspace{2mm}
(b) Relative improvements using embeddings extracted from an encoder using SincNet initialization based on the CQT filterbank\\
\begin{tabular}{l|c|cc|cccccc}
\toprule
& & \multirow{2}*{NSynth} & \multirow{2}*{RWC} & \multicolumn{2}{c}{MLP} & \multicolumn{2}{c}{SVM} & \multicolumn{2}{c}{DT}\\
\cline{5-10}
& \#classes & & & Micro F1 $\uparrow$ & Macro F1 $\uparrow$ & Micro F1 $\uparrow$ & Macro F1 $\uparrow$ & Micro F1 $\uparrow$ & Macro F1 $\uparrow$\\
\midrule
Instrument Family (NSynth) & 10 & $\surd$ & & 4.82 & 32.98 & 4.81 & 32.96 & 2.99 & 21.12  \\
Source & 3 & $\surd$ & & 1.50 & 4.23 & 1.50 & 4.23 & 0.86 & 2.87\\
Pitch & 10 & $\surd$ & & 3.42 & 19.59 & 3.59 & 19.90 & 0.97 & 7.38 \\
Velocity & 5 & $\surd$ & & 0.70 & 4.06 & 0.83 & 4.50 & 0.12 & 2.24 \\
Instrument Family (RWC) & 7 & & $\surd$ & 2.45 & 16.62 & 2.43 & 16.54 & 2.05 & 13.27  \\
Dynamics & 3 & & $\surd$ & 0.89 & 2.90 & 0.95 & 3.01 & 0.34 & 1.74 \\
Playing Style (Strings) & 20 & & $\surd$ & 2.31 & 28.48 & 2.69 & 34.53 & 0.76 & 7.86 \\
Playing Style (Woodwinds) & 7 & & $\surd$ & 1.13 & 8.83 & 1.22 & 9.40 & 0.42 & 4.13 \\
Playing Style (Brass) & 18 & & $\surd$ & 2.91 & 32.96 & 3.00 & 33.10 & 1.15 & 14.15 \\
\bottomrule
\end{tabular} \\
\label{tab:probing-relative}
\end{table*}

From the figure and table, we can first see that the embedding vectors are rich representations that can be used to predict all types of metadata labels better than majority voting, although the encoder was trained using the instrument and instrument-family categories only.

Moreover, interestingly, from the comparison of the results of the two types of embeddings, we can see that shallow classifiers using embeddings extracted from the encoder using SincNet initialization based on the CQT filters had larger relative improvements for the playing style prediction tasks. For instance, relative improvement value of MLP trained on the embeddings extracted from the encoder using SincNet initialization based on the CQT filters is 2.31 whereas that based on the Mel filters is 0.80 in terms of Micro F1. We can see the same tendency in woodwinds and brass. It seems that the use of the CQT filterbank helped to encode information relevant to the playing style, even though these labels are not used during the training phase of the instrument encoder. Therefore, we can conclude that the instrument embedding vector is enriched with the playing style information specific to the instrument family and this resulted in reduced EERs of unknown instrument variations on the RWC database.

\section{Conclusion}
\label{sec:conclusion}
We have used ASV techniques to construct and evaluate a musical instrument embedding space capable of meaningfully representing unseen instruments and instrumental variations.  We explored different training strategies for the embedding model and found through ablation experiments that data augmentation and use of angular softmax were important components that helped to improve EER for the verification of unseen instruments.  We also found that using a CQT-based front-end gave further improvements over a Mel filterbank based one, and that we could obtain EER values of less than 3\% on average.  We also studied the structure and content of the learned embedding space using t-SNE visualizations and probing tasks.  We found that including instrument family labels as a multi-task learning target helped to regularize the embedding space and incorporate structure pertaining to instrument family, which also made instrument family information more available to the downstream probing task.  The probing experiments also showed that playing style information is also contained in the embeddings and the CQT filterbank helps to encode this information, even though labels for this information were not explicitly available at training time.
    
In future work, we will continue to experiment with the configurations of the embedding model to determine whether more adjustments can be made to the parameters and components to better model musical sounds as opposed to speech, and to further improve representation of unseen instruments.  For instance, we may adjust the number of LDE dictionaries, the angular softmax margin, or the type of ResNet.  There are also many further interesting directions for analysis of what different components of the model learns, such as what kind of information each of the LDE dictionaries contains.  More importantly we plan to use these embeddings for downstream music synthesis tasks such as multi-instrument synthesis and timbre transfer, and to use these tasks to further evaluate and analyze embeddings of unseen instruments.  

\bibliographystyle{IEEEtran}
\bibliography{main}

\hfill

\begin{IEEEbiography}[{\includegraphics[width=1in,height=1.25in,clip,keepaspectratio]{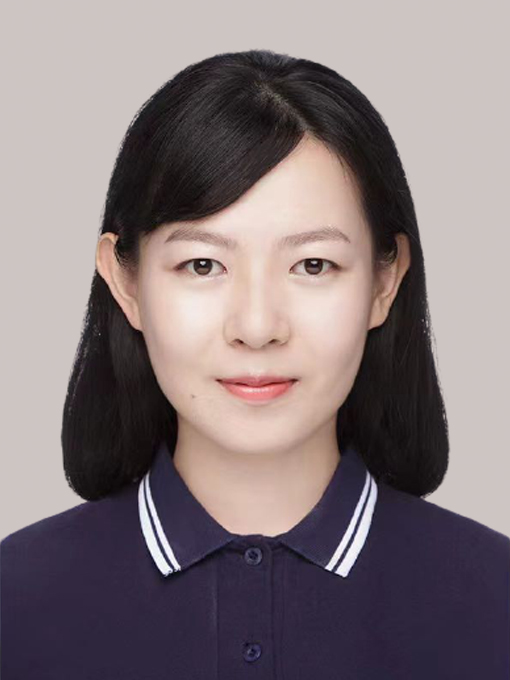}}]{Xuan Shi} received a B.Sc degree in communication engineering from Shanghai University, Shanghai, PRC, in 2019. Since 2020, she has been a master student in electrical engineering at the University of Southern California, Los Angeles, CA, USA. 
\end{IEEEbiography}

\begin{IEEEbiography}[{\includegraphics[width=1in,height=1.25in,clip,keepaspectratio]{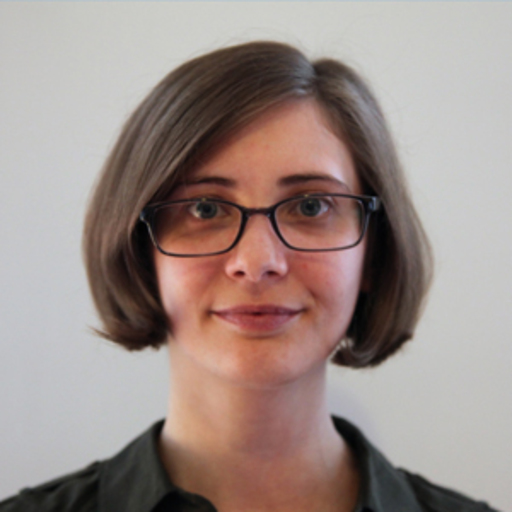}}]{Erica Cooper} (M'19) received a B.Sc. degree and M.Eng. degree both in electrical engineering and computer science from the Massachusetts Institute of Technology, Cambridge, MA, USA, in 2009 and 2010, respectively. She received a Ph.D. degree in computer science from Columbia University, New York, NY, USA, in 2019. Since 2019, she has been a Project Researcher with the National Institute of Informatics, Chiyoda, Tokyo, Japan. Her research interests include statistical machine learning and speech synthesis. Dr. Cooper's awards include the 3rd Prize in the CSAW Voice Biometrics and Speech Synthesis Competition, the Computer Science Service Award from Columbia University, and the Best Poster Award in the Speech Processing Courses in Crete.
\end{IEEEbiography}

\begin{IEEEbiography}[{\includegraphics[width=1in,height=1.25in,clip,keepaspectratio]{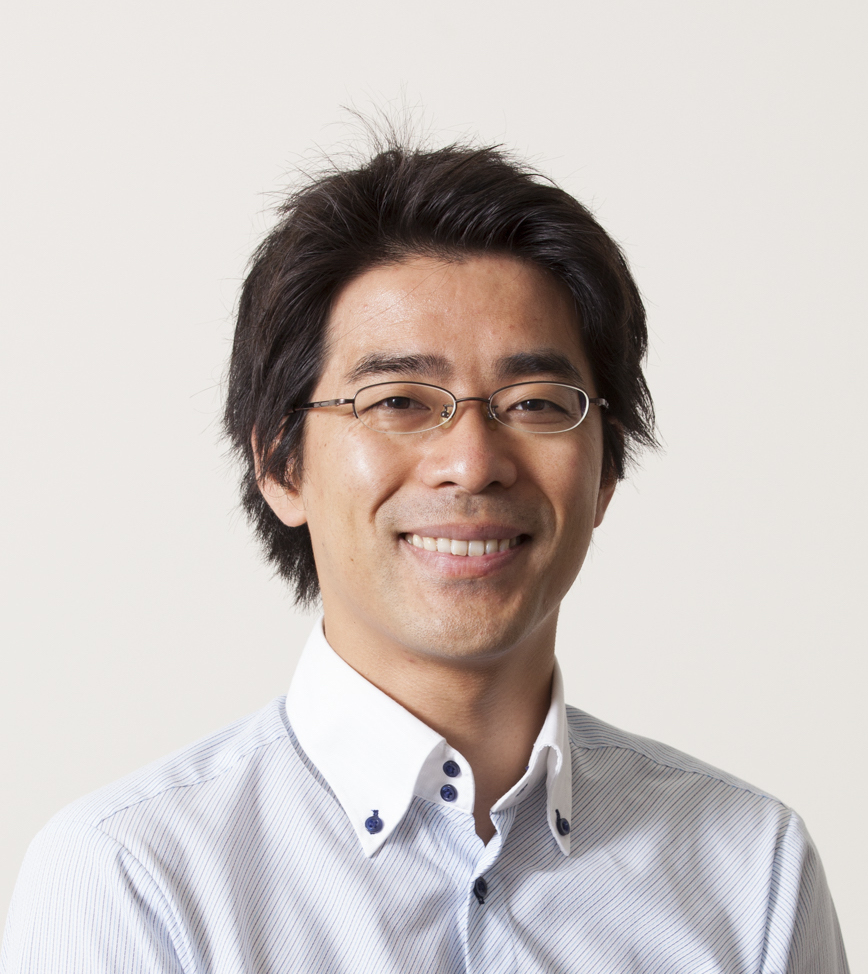}}]
{Junichi Yamagishi} (SM'13) received a Ph.D.\ degree from the Tokyo Institute of Technology (Tokyo Tech), Tokyo, Japan, in 2006. From 2007-2013, he was a research fellow in the Centre for Speech Technology Research (CSTR) at the University of Edinburgh, UK. He was appointed Associate Professor at the National Institute of Informatics, Japan, in 2013. He is currently a Professor at NII, Japan. His research topics include speech processing, machine learning, signal processing, biometrics, digital media cloning, and media forensics. 

He served previously as co-organizer for the bi-annual ASVspoof Challenge and the bi-annual Voice Conversion Challenge. He also served as a member of the IEEE Speech and Language Technical Committee (2013-2019), an Associate Editor of the IEEE/ACM Transactions on Audio Speech and Language Processing (2014-2017), and a chairperson of ISCA SynSIG (2017- 2021). He is currently a PI of JST-CREST and ANR supported VoicePersona project and a Senior Area Editor of the IEEE/ACM TASLP.
\end{IEEEbiography}

\end{document}